\documentclass[%
 aip,
 jmp,%
 amsmath,amssymb,
 reprint,%
]{revtex4-1}

\usepackage{graphicx}
\usepackage{dcolumn}
\usepackage{bm}
\usepackage{amsthm}
\usepackage{color}

\newtheorem{proposition}{Proposition}
\DeclareMathOperator\arctanh{arctanh}

\begin{document}

\title[Stokes flow in a channel with permeable boundaries]{An exact solution for Stokes flow in a channel with arbitrarily large wall permeability} 

\author{G. Herschlag}
 \affiliation{Mathematics Department, Duke University.}
  \email{gjh@math.duke.edu.}
  
\author{J.-G. Liu}%
\affiliation{Mathematics Department, Duke University.}
\affiliation{Physics Department, Duke University.}

\author{A. T. Layton}%
\affiliation{Mathematics Department, Duke University.}

\date{\today}


\begin{abstract}
We derive an exact solution for Stokes flow in a channel with permeable walls.
At the channel walls, the normal component of the fluid velocity is described by Darcy's law and the tangential component of the fluid velocity is described by the no slip condition.  
The pressure exterior to the channel is assumed to be constant. 
Although this problem has been well studied, typical studies assume that the permeability of the wall is small relative to other non-dimensional parameters; this work relaxes this assumption and explores a regime in parameter space that has not yet been well studied.
A consequence of this relaxation is that transverse velocity is no longer necessarily small when compared with the axial velocity.
We use our result to explore how existing asymptotic theories break down in the limit of large permeability for channels of small length.
\end{abstract}

\keywords{Filtration; permeable boundaries; Stokes flow}
\maketitle

\section{Introduction}

There is a great deal of interest in the analysis and simulation of fluid flow along permeable tubes, in large part owing to the multitude of applications. 
In engineering, examples include ultrafiltration membrane systems used in water treatment. 
In biology, examples include glomerular filtration in the kidney, dialysis machines, and capillary transport.
Given the wide range of applications, much effort has been dedicated to modeling and computing fluid flow along tubes or channels with porous walls. 
In a 1953 classical paper, perhaps one of the most frequently cited analytical works in this area, Berman used perturbation methods to obtain solutions to the Navier-Stokes equations that describe fluid flow in a rectangular slit with two equally porous walls. 
The flow is assumed to be laminar and the normal velocity at the wall is assumed to be a known constant {\it a priori} and independent of position \cite{Berman1953}. 
Berman's result was later extended to a cylindrical geometry by \citet{YuanFinkelstein1956}, again with constant normal velocity at the walls, which was further expounded upon by \citet{Terrill1964} and \citet{Terrill1965}.

In membrane transport, the normal fluid velocity is frequently driven by hydrostatic pressure gradient.   The classical assumption is that the outer and inner channel pressures have the same gradient, leading to the assumption of a constant normal velocity.
In practice, however, this assumption is typically violated, and much work has gone into changing the assumption on the pressure exterior of the channel to be a constant along the axial direction, rather than having a similar gradient to the flow in the channel.  
Enforcing Darcy's law at the boundaries, this change in assumptions implies that the normal wall velocity is a function of the hydrostatic pressure.
A notable study of this new boundary condition is provided by \citet{Galowin1974}, in which the flow profile for a semi-infinite pipe with a closed end wall is considered with normal velocity described by Darcy's law, and the external pressure assumed to be constant.  
This work was later expanded by \citet{Granger1988} who assume one permeable and one impermeable wall, along with parabolic inflow conditions at a certain point within the pipe.  
Indeed, much effort has been devoted toward this topic. 
More recently, Haldenwang has studied the problem in more detail; nonetheless, this work still considers a channel or pipe of finite length with prescribed inflow and outflow conditions \cite{Haldenwang2007, Haldenwang2011, Bernales2014}; the inflow condition in these last set of papers is matched with the analytic solution by \citet{Berman1953} in which the normal velocities at the wall are matched.  In the work of \citet{Tilton2012}, the authors relaxed the assumption of a finite length pipe and found a self similar profile in a pipe of arbitrary length.

The works mentioned above assume either small permeability or a small ratio between transverse and axial flow velocities.  With these assumptions, leading order asymptotic expansions result in parabolic profiles in the axial flow profile at vanishing Reynolds number that are independent of permeability.  There is, however, no analytic result to assess the correctness of this result, and it is therefore valuable to ask if and how the predicted Stokes flow of existing asymptotic studies breaks down as permeability increases. 

It has been noted throughout the literature \cite{Haldenwang2007, Haldenwang2011, Bernales2014} that as permeability grows the convective terms of the Navier-Stokes equations become important at smaller length scales.  This observation is attributable to the fact that the velocity profile grows exponentially, and the rate of this exponential growth increases with permeability.  Hence we expect an exact solution at zero Reynolds number to be physically reliable only in a finite region that will depend both on the Reynolds number and permeability of the wall.  Nevertheless, such a solution would be valuable as it would (1) provide a mechanism to determine the break down of the existing theory for zero Reynolds number and (2) provide an analytic result with which to compare numerical studies (provided that the Reynolds number remains small in the domain of interest).

Therefore, in the present work we examine channel flow in the Stokes regime in which at the channel walls the normal and tangential components of the fluid velocity are described by Darcy's law and the no slip condition, respectively.  
The pressure exterior to the channel is assumed to be uniform and similarly to \citet{Tilton2012}, we do not assume the structure of an inlet flow and allow such structure to arise from the equation solution.  Also similar to much of the existing work in this field, we assume a symmetric flow profile in the transverse direction (see for example Refs. \cite{Haldenwang2007, Haldenwang2011, Tilton2012,Bernales2014}).
These assumptions are consistent with previous problem statements in this field, and for biological flows the porosity of the channels will be due to orthogonal water channels making the no slip condition in the axial velocity more robust.  We determine an exact solution for this problem, compare our result with the existing theory, and then numerically validate our result in a finite domain of interest.


%

Although we have limited our study of high permeability to only consider the zero Reynolds number regime, we note that this choice is well within the confines of many applications in this field and, in particular, many of the biological applications listed above occur at negligible Reynolds number.  In \citet{Pozrikidis2010}, for example, the authors have analyzed a similar problem in modeling capillary blood flow in a numerical study, in which the author assumes a region of impermeable pipe feeds into a permeable region and the flow is described by Stokes equations.  
Furthermore, fluid flow in the collecting ducts of rat kidneys also occur at small Reynolds number, occurring at most on the order of $10^{-4}$ (see Refs. \onlinecite{Pennell1974,Knepper1977}).
Thus an exact solution at zero Reynolds number is a valuable contribution to this field and may be used as a comparison point in future asymptotic theory.  

In section \ref{sec:solution}, we present and solve the system of equations described above. 
In section \ref{sec:smallperm}, we then demonstrate the break down of existing asymptotic theories for small permeability as the permeability grows.
In section \ref{sec:numerical} we demonstrate that there are regions of the parameter space in which our exact solution is significantly more accurate than the existing theory even when inertial effects are considered.
We conclude with a discussion in section \ref{sec:disc}, and suggest a physical experiment that may be preformed to validate our analysis.

The solution, presented in section \ref{sec:solution}, is computed using a non-standard approach and thus we outline our method here.  First we non-dimensionalize the system, but keep two free variables in the non-dimensionalization.  We next take an ansatz that the pressure will satisfy a Robin boundary condition; keeping the two variables free in the non-dimensionalization allows us to scale this Robin condition in such a way that incompressibility will be satisfied.  
The ansatz allows us to decouple the system and first determine the pressure.  
We then use the pressure to find the velocity profile and finally use the incompressibility constraint to restrict the two free parameters found in the non-dimensionalization.





\section{Stokes flow solution to channel with permeable boundaries}\label{sec:solution}
We examine the Stokes flow equation for a channel with arbitrarily large permeability at the boundaries.  
We assume that the pressure outside of the channel is a constant $P_{o}$, and without loss of generality, set this value to be zero.  Let the channel walls be separated by a distance $2r$.  Let $x$ describe the location parallel to the axial position and $y$ describe the location in the transverse direction.  The flow in each direction is described by $u(x,y)$ and $v(x,y)$, respectively.
We assume that at position $x=x_0$, the central pressure gradient $\beta$ and the inner channel pressure at the wall are given by
\begin{eqnarray}
p(x_0,\pm r) = P_{tm}, \quad \beta = p_x(x_0,0). \label{eqn:osys1}
\end{eqnarray}
Without loss of generality we set $x_0=0$. 
We will show below a one-to-one correspondence between $\beta$ and the average flow profile:
\begin{equation}
\bar{U} = \frac{1}{2r}\int_{-r}^r u(x_0,y)dy.
\end{equation}
Although the latter is a more standard choice (see for example \citet{Tilton2012}) we elect to use the former for mathematical convenience.

Fluid flow between the interior and exterior of the channel is assumed to be driven by the pressure difference across the two regions.  In the physical case of filtration Darcy's law is typically assumed in the normal direction at the wall and a no slip condition is assumed in the tangential direction (see, for example, Refs. \onlinecite{Regirer60, Karode2001,Tilton2012, Bernales2014}).  We also use these boundary conditions on the channel walls, using Darcy's law with coefficient $\kappa=k/\mu h$ in the normal direction, where $k$ is the permeability, $\mu$ is the viscosity, and $h$ is the width of the channel connecting the outer and inner fluids.  No slip conditions are used for the tangential component of the boundary.
For the two dimensional setting, the Stokes equations may then be written as
\begin{eqnarray}
&&p_x = \mu\Delta u, \quad p_y = \mu\Delta v,\\
&&0=u_x+v_y,\\
&&u(x,\pm r) = 0,\quad v(x,\pm r) = \pm \kappa p,\label{eqn:osyse}
\end{eqnarray}
where $\mu$ is the dynamic viscosity, and $u$ and $v$ are the velocity components in the $x$ and $y$ directions, respectively.  
We note that this system of equation matches the zeroth order term in an asymptotic expansion of the Navier-Stokes equations about small Reynolds number (see Ref. \onlinecite{Regirer60}).



\subsection{The non-dimensional system}
To derive a solution, we first non-dimensionalize and identify the important non-dimensional parameter.  We rescale by taking
\begin{eqnarray}
x&=& \frac{r}{\gamma}\tilde{x},\quad
y=\frac{r}{\gamma} \tilde{y}, \quad
p = \frac{\beta r}{\gamma} \tilde{p},\\
u &=& \beta r^{\alpha} \gamma^{-\alpha}\kappa^{2-\alpha} \mu^{1-\alpha}\tilde{u},\\
v &=&  \beta r^{\alpha} \gamma^{-\alpha}\kappa^{2-\alpha} \mu^{1-\alpha} \tilde{v},
\end{eqnarray}
where $\gamma,\alpha\in\mathbb{R}$ with $\gamma>0$.  Substituting the non-dimensional rescaling into the original system and dropping the tildes for convenience, we obtain
\begin{eqnarray}
&&p(0,\pm \gamma) = \mathcal{B}, \quad  p_x(0,0)=1,\label{eqn:nondimvel}\\
&&p_x = \mathcal{A}^{2-\alpha}\Delta u,\quad p_y = \mathcal{A}^{2-\alpha}\Delta v,\\
&&0=u_x+v_y,\label{eqn:divfree}\\
&&u(x,\pm \gamma) = 0,\quad v(x,\pm \gamma) = \pm \mathcal{A}^{\alpha-1} p,\label{eqn:nondimbdry}
\end{eqnarray}
where $\mathcal{A} = \gamma\mu\kappa/r$ and $\mathcal{B}={P_{tm} \gamma}/{\beta r}$.  We define a special case where $\gamma=1$ as $\mathcal{A}_1 = \mu\kappa/r$ and $\mathcal{B}_1= {P_{tm}}/{\beta r}$ which will be used below. 

It is not standard to introduce $\gamma$ and $\alpha$ and values for these parameters are typically chosen implicitly along with the non-dimensionalization, the standard being $\gamma=1$ and $\alpha=2$.  
We note that the reason for leaving $\alpha$ free at the moment is that we will take an ansatz that requires the ability to allow a ratio of arbitrary size between $\mathcal{A}^{2-\alpha}$ and $\mathcal{A}^{\alpha-1}$, that is we will require the ability to freely adjust the value $\mathcal{A}^{3-2\alpha}$ as a function of $\alpha$.  The ability to scale this quantity may only be achieved if $\mathcal{A}\neq 1$ and thus we introduce a second scaling parameter $\gamma$ to rescale the non-dimensional length.  We note then that if $\mathcal{A}=1$ for a given value of $\gamma$, we may simply assign a different value to $\gamma$ which will ensure $\mathcal{A}\neq 1$.  Noting that there is a potential degeneracy in the scaling with $\gamma$ fixed and $\alpha$ free, it is reasonable to propose that we instead fix $\alpha$ and leave $\gamma$ free.  Although this is possible, $\emph{a priori}$ it is unclear which value we should assign to $\alpha$. Below we will show that $\alpha=1$ is the proper choice, which is different from the standard choice of $\alpha=2$.  

\subsection{Establishing the ansatz}
Equations \ref{eqn:nondimvel}-\ref{eqn:nondimbdry} form a complex and coupled system.  We may, however, attempt to decouple the pressure term from the velocity equations by setting $\bar{v}=v+p_y$ and assume $\bar{v}$ vanishes at the boundaries.  This assumption requires that pressure satisfy a Robin boundary condition and the equations may be rewritten as
\begin{eqnarray}
&&p(0,\pm \gamma) = \mathcal{B}, \quad  p_x(0,0)=1,\label{eqn:nondimvel2a}\\
&&\Delta p = 0, \quad \partial_{y}p(x,\pm\gamma)=-\mathcal{A}^{\alpha-1}p(x,\pm\gamma)\label{eqn:nonppoisson}\\
&&p_x = \mathcal{A}^{2-\alpha}\Delta u,\quad p_y = \mathcal{A}^{2-\alpha}\Delta \bar{v},\label{eqn:nondimvel2b}\\
&&0=u_x+\bar{v}_y-p_{yy},\label{eqn:divfree2}\\
&&u(x,\pm \gamma) = 0,\quad \bar{v}(x,\pm \gamma) = 0,\label{eqn:nondimbdry2}
\end{eqnarray}
which appears to be an overdetermined system.  We note that a solution to this system will provide a solution to the original set of equations \ref{eqn:nondimvel}-\ref{eqn:nondimbdry}.  To see this we note the boundary conditions at the channel walls for $p$ and $v$ are equivalent via equations \ref{eqn:nondimbdry} and \ref{eqn:nonppoisson} which imply that
\begin{equation}
v(x,\pm \gamma)=\mp \partial_y p(x,\pm\gamma)=\pm \mathcal{A}^{\alpha-1}p(x,\pm\gamma).
\end{equation}
The rest of the relationships are straightforward.  A solution to the system given by equations \ref{eqn:nonppoisson}-\ref{eqn:nondimbdry2} provides a solution to the system given by equations \ref{eqn:nondimvel}-\ref{eqn:nondimbdry}.   In attempting to solve the new system we note that the equation for pressure is decoupled and thus we may solve it without knowing the velocity profile.  After solving for the pressure we may then solve the Poisson equations for the velocity $u$ and the adjusted velocity $\bar{v}$.  It is unclear however that equation \ref{eqn:divfree2} will be satisfied by this attempt to find a solution. We note that we will find $u$ and $\bar{v}$ to be inversely proportional to $\mathcal{A}^{2-\alpha}$.  Having left both $\gamma$ and $\alpha$ free, we are free to scale these solutions so that $u_x+\bar{v}_y$ has the possibility to become proportional to $p_{yy}$.  Although it is not obvious at the outset, incompressibility will be shown to be satisfied with the correct scaling of $\alpha$ and $\gamma$.

In addition to the ansatz of a Robin boundary condition for pressure (and that the resulting equations will remain incompressible), we also assume an axisymmetric flow profile so that that $v$ is odd in $y$ and $u$ and $p$ are even in $y$.

For the remainder of this section we show that the new problem statement allows us to generate solutions for the Stokes equations for all values of $\mathcal{A}$ except for $\mathcal{A}=1$.  There is one exception to this statement in which solutions are defined for all values of $\mathcal{A}$ at a special value of $\gamma$.  We will show that the special choice of $\gamma$ results in setting $\alpha=1$.




\subsection{Solving the system}
We begin by solving for $p$ using normal mode analysis.  Due to the symmetry assumption that $p$ is even in $y$, the pressure can be written as
\begin{equation}
p(x,y) = \sum_{n=0}^\infty \hat{p}_n(x) \cos(\lambda_n y),
\end{equation}
with eigenvalues satisfying
\begin{equation}
\lambda_n \sin(\lambda_n \gamma) = \mathcal{A}^{\alpha-1} \cos(\lambda_n \gamma).
\label{eqn:eigenmodes_og}
\end{equation}
We note, and will later use, that equation \ref{eqn:eigenmodes_og} implies
\begin{equation}
\pi (n+1/2)/\gamma > \lambda_n > \pi n/\gamma ,
\label{eqn:eigbound}
\end{equation}
for $\mathcal{A} > 0$.

Substituting into the equation \ref{eqn:nonppoisson} and solving, we find
\begin{equation}
p(x,y) = \sum_{n=0}^\infty \left(c_n \sinh(\lambda_n x)+d_n \cosh(\lambda_n x)\right) \cos(\lambda_n y).
\label{eqn:pressuregeneric}
\end{equation}

The unknowns $c_n$ and $b_n$ satisfy
\begin{eqnarray}
&&p_x(0,0) =  \sum_{n=0}^\infty c_n = 1 \\
&&p(0,\pm \gamma) =  \sum_{n=0}^\infty d_n \cos(\lambda_n \gamma) = \mathcal{B} \\
\end{eqnarray}
We note that in the case that $\kappa=\mathcal{A}=0$, we have $\lambda_0=0$ and the solution for pressure will change to
\begin{eqnarray}
p(x,y) &=& c_0 x +\sum_{n=1}^\infty c_n \sinh(\lambda_n x) \cos(\lambda_n y)\nonumber \\
&& +d_0 +\sum_{n=1}^\infty d_n \cosh(\lambda_n x) \cos(\lambda_n y).
\end{eqnarray}
The typical assumption that leads to Poiseuille flow is that $c_n=0$ for $n>0$ which ensures a linear growth in the pressure as $x$ goes to $\pm\infty$.  Indeed, we can see that as $\kappa$ approaches 0 from above, $\lambda_n$ approaches $\pi n/\gamma$ from the right.  

In the limit of $\kappa\rightarrow0$ we expect to locally recover Poiseuille flow, meaning that for small values of $\kappa$ the pressure should be linear about a growing neighborhood of $x$.  To achieve this we must have that for all $n>0$, $c_n\rightarrow 0$ and $d_n\rightarrow 0$ as $\kappa\rightarrow 0$.  Although it may be possible to have higher order modes appear in a formal solution for the pressure, here we focus on the zeroth order mode and assume that $c_n=0$ and $d_n=0$ for $n>0$.  We will see below that this assumption is necessary for our theory to be consistent with the existing theory in the literature.

We therefore write an expression for the non-dimensional pressure to be
\begin{equation}
p(x,y) =\left\{
\begin{array}{ll}
      x + \mathcal{B} , &\lambda_0=0 \\[2pt]
      \frac{1}{\lambda_0}\sinh(\lambda_0 x)\cos(\lambda_0 y)&\\
      \phantom{asdf}+\frac{\mathcal{B}}{\cos(\lambda_0 \gamma)}\cos(\lambda_0 y), & \lambda_0>0.
    \end{array} \right.
\label{eqn:nondimpress}
\end{equation}
We note that the exponential growth of the pressure in the axial direction is well known and we compare growth coefficients to earlier work in section \ref{sec:compare}.  

We now have a simple equation for $u$ and $\bar{v}$ for $\kappa\neq0$
\begin{eqnarray}
\left(\cosh(\lambda_0 x) +\frac{\mathcal{B} \lambda_0\sinh(\lambda_0 x)}{\cos(\lambda_0 \gamma)}\right)  &=& \frac{\mathcal{A}^{2-\alpha} \Delta u}{\cos(\lambda_0 y)},\label{eqn:pu}\\
-\left(\sinh(\lambda_0 x)+\frac{\mathcal{B}\lambda_0\cosh(\lambda_0 x)}{\cos(\lambda_0 \gamma)}\right) &=& \frac{\mathcal{A}^{2-\alpha} \Delta \bar{v}}{\sin(\lambda_0 y)}\label{eqn:pv},
\end{eqnarray}
where $\bar{v} = v+p_y$ with boundary conditions located at $\pm\gamma$ and given in equation \ref{eqn:nondimbdry2}.  
We use normal mode analysis to solve both equations below.

\subsubsection{Solution for $u$}
As we have assumed $u$ to be even in $y$, we substitute a Fourier series of the form 
\begin{eqnarray}
\sum_{n=0}^\infty \hat{u}_n(x)\cos(\omega_n y),\quad \omega_n = \left(\frac{1}{2}+n\right)\pi/\gamma,
\end{eqnarray}
into equation \ref{eqn:pu} to solve the Poisson equation.  
Solving the orthogonal mode equations gives
\begin{eqnarray}
\hat{u}_n(x) &=& -\frac{d_n \left(\cosh(\lambda_0 x)+\frac{\mathcal{B} \lambda_0}{\cos(\lambda_0 \gamma)}\sinh(\lambda_0 x)\right)}{\mathcal{A}^{2-\alpha} (\omega_n^2-\lambda_0^2)}\nonumber\\
&&+a^u_n\cosh(\omega_n x)+b^u_n\sinh(\omega_n x),
\label{eqn:usol}
\end{eqnarray}
where the $a^u_n$'s and $b^u_n$'s are unknowns and are solutions to the homogenous flow equations where $\beta=P_{tm}=0$.  
The coefficients $d_n$ arise from Fourier expanding $\cos(\lambda_0 y)$ given by
\begin{eqnarray}
d_n &=& \frac{\int_{-\gamma}^\gamma \cos(\omega_n y) \cos(\lambda_0 y) dy}{ \int_{-\gamma}^\gamma \cos(\omega_n y)^2 dy}= \frac{2 (-1)^n \omega_n \cos(\lambda_0 \gamma)}{\gamma(\omega_n^2-\lambda_0^2)}.
\end{eqnarray}
We expect that there is no flow when the pressure gradient $\beta$ and the pressure across the channel $\mathcal{B}$ ($\propto P_{tm}$) go to zero meaning that the homogenous solutions are zero and $a^{u}_n=b^{u}_n=0$ for all $n$ in this limit.  The non-dimensional parameter $\mathcal{A}$ does not depend either on $\beta$ or $P_{tm}$ and thus we set $a^{u}_n=b^{u}_n=0$.

\subsubsection{Solution for $\bar{v}$}
As we have assumed $v$ to be even in $y$, we expand it in a Fourier series of the form 
\begin{eqnarray}
\sum_{n=0}^\infty \hat{v}_n(x)\sin(\bar{\omega}_n y),\quad \bar{\omega}_n &=&  n \pi/\gamma,
\end{eqnarray}
into equation \ref{eqn:pv} to solve the Poisson equation.  
Solving the orthogonal mode equations gives
\begin{eqnarray}
\hat{\bar{v}}_n(x) &=& \frac{\bar{d}_n \left(\sinh(\lambda_0 x)+\frac{\mathcal{B} \lambda_0}{\cos(\lambda_0 \gamma)}\cosh(\lambda_0 x)\right)}{\mathcal{A}^{2-\alpha} (\bar{\omega}_n^2-\lambda_0^2)}\nonumber\\&&+a^v_n\sinh(\bar{\omega}_n x)+b^v_n\sinh(\bar{\omega}_n x),
\label{eqn:vbarsol}
\end{eqnarray}
where the $a^v_n$'s and $b^v_n$'s are unknowns that are solutions to the homogenous flow equations where $\beta=P_{tm}=0$. 
The coefficients $\bar{d}_n$ arise from Fourier expanding $\cos(\lambda_0 y)$ given by
\begin{eqnarray}
\bar{d}_n &=& \frac{\int_{-\gamma}^\gamma \sin(\bar{\omega}_n y) \sin(\lambda_0 y) dy}{ \int_{-\gamma}^\gamma \sin(\bar{\omega}_n y)^2 dy}= \frac{2(-1)^{n+1} \bar{\omega}_n \sin(\lambda_0\gamma)}{\gamma(\bar{\omega}_n^2+\lambda_0^2)}.
\end{eqnarray}

Similarly as for $u$, we set $a^{v}_n=b^{v}_n=0$ for all $n$.

\subsubsection{Enforcing incompressibility}
To enforce that equations \ref{eqn:usol} and \ref{eqn:vbarsol} provide a solution to the original non-dimensional equations \ref{eqn:nondimvel} and \ref{eqn:divfree}, we must check that the flow field is divergence free given a proper choice of $\alpha$ and $\gamma$.  Taking partial derivatives, we find
\begin{widetext}
\begin{eqnarray}
u_x &=& -\frac{\left(\lambda_0 \sinh(\lambda_0 x)+\frac{\mathcal{B} \lambda_0^2}{\cos(\lambda_0 \gamma)}\cosh(\lambda_0 x)\right)}{\mathcal{A}^{2-\alpha}}\sum_{n=0}^\infty\frac{d_n}{\omega_n^2-\lambda_0^2} \cos(\omega_n y),\\
v_y &=& \frac{ \left(\lambda_0 \sinh(\lambda_0 x)+\frac{\mathcal{B} \lambda_0^2}{\cos(\lambda_0 \gamma)}\cosh(\lambda_0 x)\right)}{\mathcal{A}^{2-\alpha}}\sum_{n=0}^\infty\frac{\bar{d}_n \bar{\omega}_n}{\bar{\omega}_n^2-\lambda_0^2} \cos(\bar{\omega}_n y)\\
&&+\left(\lambda_0\sinh(\lambda_0 x)+\frac{\mathcal{B} \lambda_0^2}{\cos(\lambda_0 \gamma)}\cosh(\lambda_0 x)\right)\cos(\lambda_0 y).
\end{eqnarray}
\end{widetext}
Incompressibility requires that
\begin{equation}
\mathcal{A}^{2-\alpha}=\sum_{n=0}^\infty\left(\frac{d_n}{\omega_n^2-\lambda_0^2} \frac{\cos(\omega_n y)}{\cos(\lambda_0 y)}-\frac{\bar{d}_n \bar{\omega}_n}{\lambda_0 (\bar{\omega}_n^2-\lambda_0^2)} \frac{\cos(\bar{\omega}_n y)}{\cos(\lambda_0 y)}\right),
\label{eqn:alphasimp}
\end{equation}
which will only have a solution if the righthand side of the equation has no dependence on $y$.  This condition is shown to be satisfied in proposition \ref{prop:yind} in appendix \ref{appA}.  In the proposition we also simplify equation \ref{eqn:alphasimp} to read
\begin{eqnarray}
\mathcal{A}^{2-\alpha}=\frac{1}{2}\left(\frac{\gamma}{\sin(\lambda_0\gamma)\cos(\lambda_0\gamma)\lambda_0}-\frac{1}{\lambda_0^2}\right)\equiv C(\lambda_0),
\label{eqn:ctoA}
\end{eqnarray}
where we have used equation \ref{eqn:clamb}.  We can further simplify this expression by dividing equation \ref{eqn:ctoA} with equation \ref{eqn:eigenmodes_og} to eliminate $\alpha$ and relate $\mathcal{A}$ to $\lambda_0$, which yields
\begin{equation}
\mathcal{A} = \frac{1}{2}\left(\frac{\gamma}{\cos^2(\lambda_0\gamma)}-\frac{\sin(\lambda_0\gamma)}{\lambda_0\cos(\lambda_0\gamma)}\right).
\label{eqn:Avslam}
\end{equation}
The righthand side of this equation is positive and increasing for $\lambda_0\in(0,\pi (2\gamma)^{-1})$ and has range $(0,\infty)$ (see proposition \ref{prop:posmono} in the appendix and figure \ref{fig:agvl}).  Therefore there is a one-to-one correspondence between $\lambda_0\in(0,\pi (2\gamma)^{-1})$ and $\mathcal{A}\in(0,\infty)$.  To find the necessary value of $\alpha$ to enforce incompressibility, we substitute the above expression for $\mathcal{A}$ into equation \ref{eqn:eigenmodes_og} which leads to an expression for $\alpha$:
\begin{equation}
\alpha = \frac{\log\left(\lambda_0 \tan(\lambda_0\gamma)\right)}{\log(\mathcal{A})}+1
\label{eqn:alphavlam}
\end{equation}
Although we have found a solution to the original non-dimensional problem (equations \ref{eqn:nondimvel}-\ref{eqn:nondimbdry}), there is a potential singularity in $\alpha$ for $\mathcal{A}=1$.  This should not be surprising as we lose the ability to scale the lefthand side of equation \ref{eqn:ctoA} in this case.  We next simplify $\alpha$ to eliminate the potential singularity.

\begin{figure}
\begin{center}
\includegraphics[height=6cm]{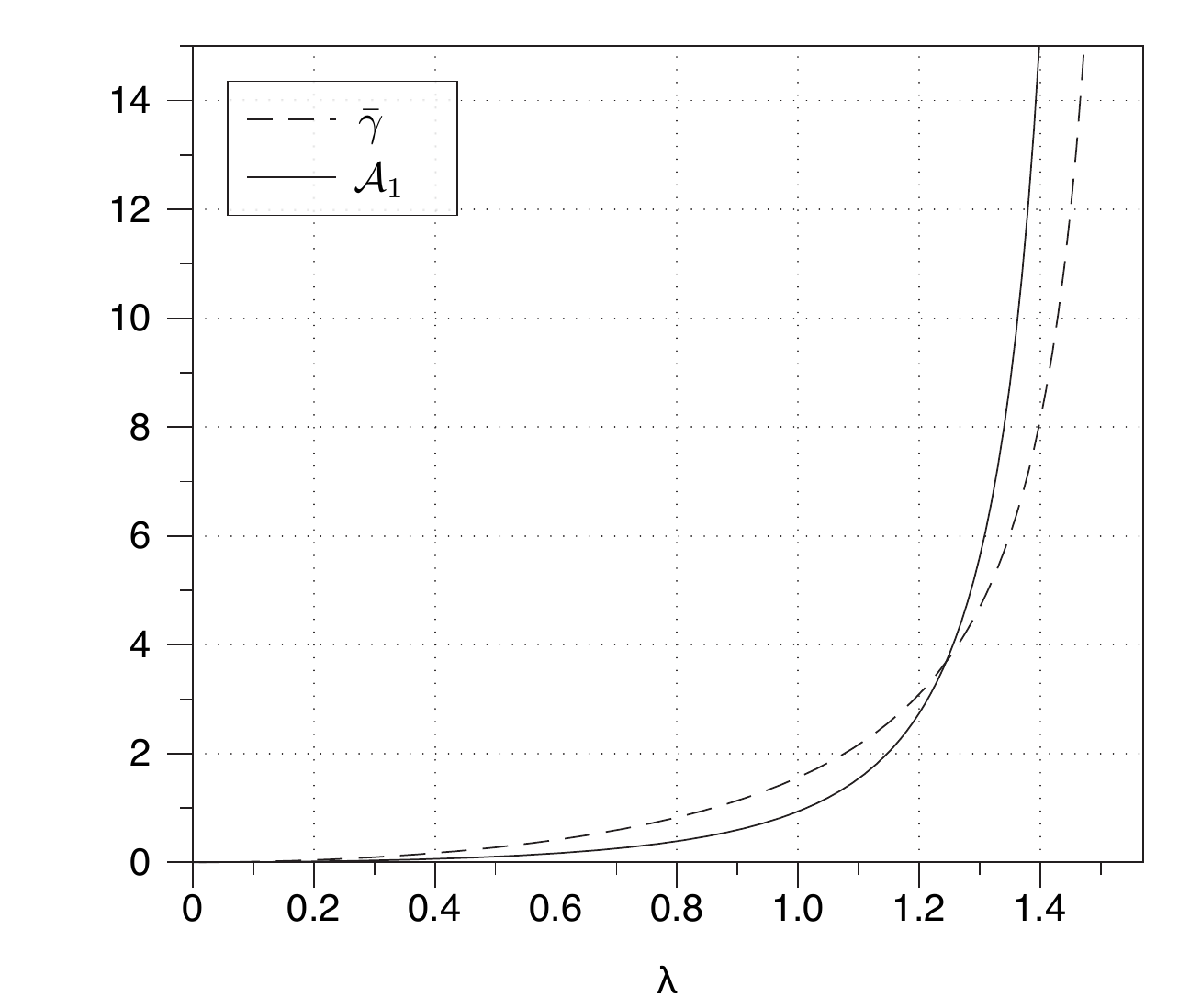}
\end{center}
\caption{$\mathcal{A}_1$ and $\bar{\gamma}$ as functions of $\lambda$.  Note that there is a bijection between $\mathcal{A}_1$ and $\lambda$ and between $\bar{\gamma}$ and $\lambda$.}
\label{fig:agvl}
\end{figure}

\subsubsection{Simplifying $\alpha$}
We begin by letting $\lambda=\lambda_0\gamma$ so that $\lambda\in(0,\pi/2)$, and will use $\mathcal{A}_1 = \mathcal{A}/\gamma = \mu\kappa/r$, as it is defined above.  Equation \ref{eqn:eigenmodes_og} may then be rewritten as
\begin{equation}
\lambda \sin(\lambda) = \mathcal{A}_1^{\alpha-1} \gamma^\alpha \cos(\lambda),
\end{equation} 
equation \ref{eqn:Avslam} as
\begin{equation}
\mathcal{A}_1 = \frac{1}{2}\left(\frac{1}{\cos^2(\lambda)}-\frac{\sin(\lambda)}{\lambda\cos(\lambda)}\right),
\label{eqn:A1vlam}
\end{equation}
and equation \ref{eqn:alphavlam} as
\begin{equation}
\alpha = \frac{\log\left(\lambda \tan(\lambda)\right) +\log(\mathcal{A}_1)}{\log(\gamma)+\log(\mathcal{A}_1)},
\label{eqn:alpha1vlam}
\end{equation}
from which we can derive an equation for $\gamma$ in terms of $\alpha$ as
\begin{equation}
\gamma = \left(\lambda \tan(\lambda)\mathcal{A}_1^{1-\alpha}\right)^{1/\alpha}.\label{eqn:uglygam}
\end{equation}

Equation \ref{eqn:A1vlam} demonstrates that the rate of pressure gain per unit channel length will be independent of $\gamma$, as expected.  Equation \ref{eqn:alpha1vlam} demonstrates that there will be an essential singularity in $\alpha$ located at $\gamma=1/\mathcal{A}_1$ (corresponding to $\mathcal{A} = 1$).  A natural way to handle this issue is to set
\begin{equation}
\alpha=1,
\end{equation}
which implies
\begin{equation}
\gamma = \lambda \tan(\lambda),
\label{eqn:gsimpvlam}
\end{equation}
and avoids the issue of the singularity within the non-dimensionalized problem.  We denote this special value of $\gamma$
\begin{equation}
\bar{\gamma}\equiv\lambda \tan(\lambda).
\end{equation}
We remark again that it is not known \emph{a priori} that $\alpha=1$ leads to a convenient non-dimensionalization and have thus left it free until this point.  We further remark that we could have chosen $\alpha$ to be any other fixed value and obtain the expression form equation \ref{eqn:uglygam}; however this equation also makes it clear that $\alpha=1$ is the aesthetically pleasing choice. This choice would not have been obvious had we chosen $\alpha$ from the start.
We plot $\mathcal{A}_1$ and $\bar{\gamma}$ versus $\lambda$ in figure \ref{fig:agvl}.

Taking into account the above simplifications, we rescale $x$ and $y$ by $\bar{\gamma}^{-1}$ and write the full non-dimensional solution in terms of $\mathcal{B}_1$ (defined above as $\mathcal{B}_1=P_{tm}/\beta r$),$\lambda$, $x$ and $y$ as
\begin{widetext}
\begin{eqnarray}
p(x,y) &=& \left(\tan(\lambda) \sinh(\lambda x)+\frac{\mathcal{B}_1 \lambda \tan(\lambda)}{\cos(\lambda)}\cosh(\lambda x) \right)\cos(\lambda y),\label{eqn:psol}\\
u(x,y) &=& \frac{32 \lambda^2 \sin(\lambda) \left(\cosh(\lambda x)+\frac{\mathcal{B}_1\lambda}{\cos(\lambda)}\sinh(\lambda x)\right)}{\pi^3 \left(\lambda \sec^2(\lambda)-\tan(\lambda)\right)} \sum_{n=0}^\infty \frac{(-1)^{n+1}(1+2n)}{\left((1+2n)^2-\left(\frac{2 \lambda}{\pi}\right)^2\right)^2} \cos\left(\frac{(1+2n)\pi}{2} y\right),\label{eqn:sol1i}\\
v(x,y) &=& \frac{4\lambda^2\cos(\lambda)\sin^2(\lambda)\left(\sinh(\lambda x)+\frac{\mathcal{B}_1\lambda}{\cos(\lambda)}\cosh(\lambda x)\right)}{\pi^3\left(\lambda- \sin(\lambda) \cos(\lambda)\right)}\sum_{n=1}^\infty \frac{(-1)^{n+1} n}{(n^2-\left(\frac{\lambda}{\pi}\right)^2)^2}\sin\left(\pi n y\right)\nonumber\\
&&+\left(\sinh(\lambda x)+\frac{\mathcal{B}_1\lambda}{\cos(\lambda)}\cosh(\lambda x)\right) \sin(\lambda y).\label{eqn:vsol} \label{eqn:sol1f}
\end{eqnarray} 
\end{widetext}

\begin{figure}
\includegraphics[width=8cm]{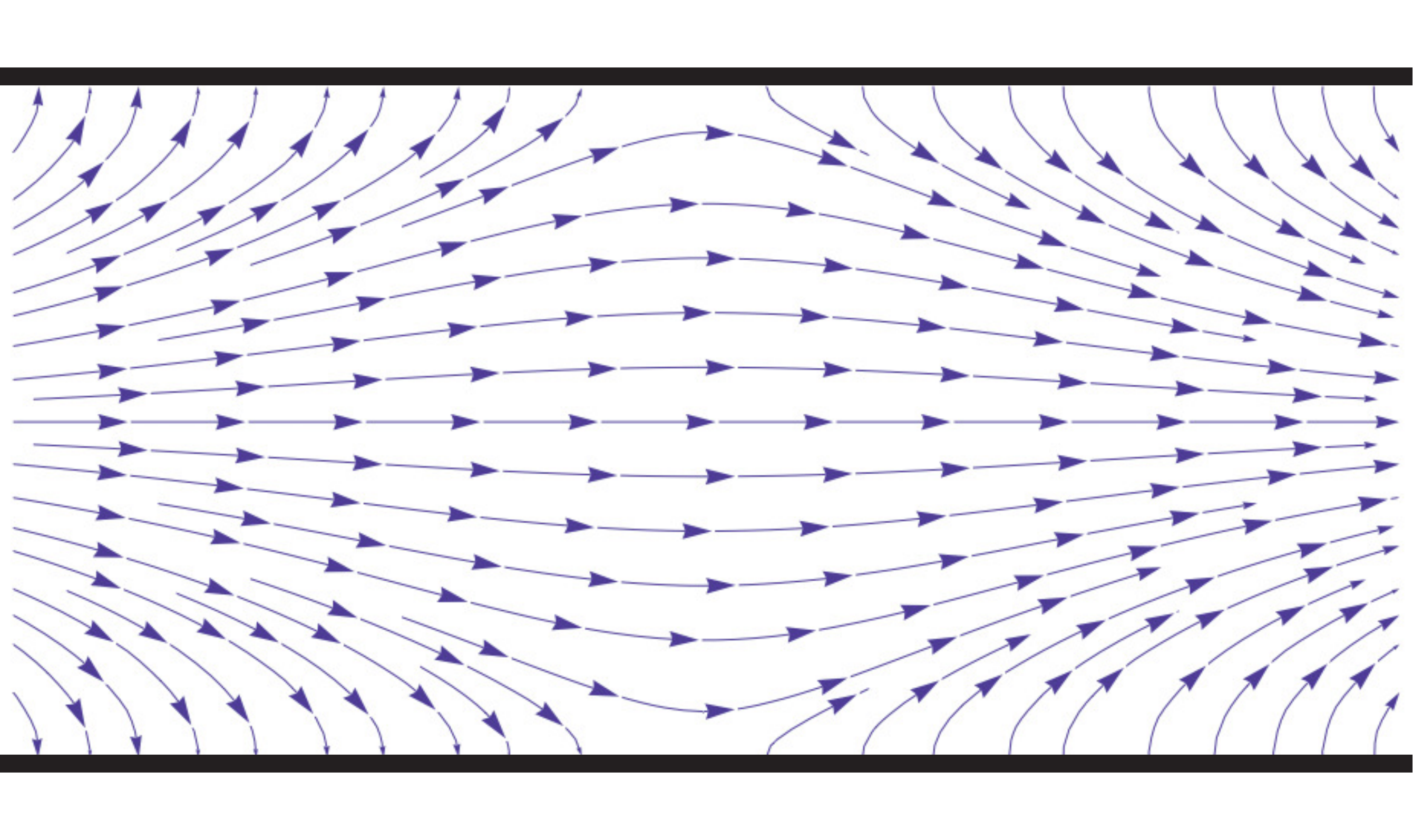}
\caption{Streamlines of channel flow for the rescaled solution presented in equations \ref{eqn:sol1i}-\ref{eqn:sol1f} in the domain $(x,y)\in[-2,2]\times[-1,1]$; in this system $\lambda=\pi/4$ (corresponding to $\mathcal{A}_1=0.36338$) and $\mathcal{B}_1=0$.}
\label{fig:streamlinesandsetup}
\end{figure}

The domain of the rescaled $y$ is now $(-1,1)$. 
 We remark again that we have a bijection between $\mathcal{A}_1$ and $\lambda$ through a transcendental equation so that knowing $\mathcal{A}_1$ will provide us with the correct choice for $\lambda$.  This solution represents a closed form solution for Stokes flow through a channel with uniformly permeable walls but having arbitrarily large permeability.  This is the major result of the paper in that all other results presented below are simple corollaries that arise from it.  We demonstrate the streamlines of this solution for $\lambda=\pi/4$ with $\mathcal{B}_1=0$ in figure \ref{fig:streamlinesandsetup}.  

As a tool for predicting fluid flow, our result is exact within the Stokes regime.  We note that in nature and physical application, permeability (i.e. $\mathcal{A}_1$ or $\lambda$) is typically small and there has been a great deal of asymptotic work done in the parameter regime for small permeability.  The present solution provides an analytic tool to determine the error in the asymptotic theory.  We also note that our solution contains exponential increase in the axial direction for both the velocity and pressure profiles, which is consistent with known results (see for example \onlinecite{Karode2001,Haldenwang2007,Tilton2012}).  This means that convective terms will become important over finite length scales; our solution should then be understood as an accurate approximation of the Navier-Stokes equation within some finite region of a channel.  We do not, however, have a prediction for which approximation to the Navier-Stokes is more accurate depending on the length of the channel, $\mathcal{A}_1$ and $\mathcal{B}_1$, so we examine this numerically below in section \ref{sec:numerical}.

Finally we remark on the relationship between the average flow velocity through $x=0$ and the point gradient of the pressure $\beta$.  Given $\mathcal{A}_1$ (and hence $\lambda$), the dimensionalized flow profile at $x=0$ may be written as 
\begin{equation}
u(0,y) = \frac{\beta r \kappa}{\lambda \tan(\lambda)} f(y;\lambda),
\end{equation}
where $f$ is the non-dimensionalized flow profile for $u$ at $x=0$ and $y\in[-r,r]$ is now back to dimensional units.  We note, and will give evidence below, that the flow profile is sign definite for all values of $\lambda$, and thus $f$ averaged over $y$ from $[-r,r]$ is non-zero.  This implies a one-to-one correspondence between the average flow profile across the $x=0$ plane and the pressure gradient at $(0,0)$ given by 
\begin{equation}
\bar{U} = -\frac{\beta \kappa}{2 \lambda \tan(\lambda)} \int_{-r}^r f(y;\lambda) dy,\label{eqn:onetoonebetaubar}
\end{equation}
for all $\kappa,\mu>0$ (and hence $\lambda\in(0,\pi/2)$). 
The above results are used below to demonstrate how asymptotic theories for small permeability developed in the literature break down as this parameter can no longer be considered small.
\section{The break down of asymptotic approximations at large permeability}\label{sec:smallperm}

We begin by analyzing the break down of the axial pressure profile, continue by analyzing the consequence of large permeability on the transverse profiles in $u$ and $v$, and conclude by discussing the error associated with the non-dimensional transition between crossflow reversal and axial flow exhaustion as well as analyze the error associated with the predicted location at which these behaviors occur.

Throughout this section we will use the leading order behavior of $\lambda$ as a function of $\mathcal{A}_1$ which may be found by noting that as permeability decreases $\mathcal{A}_1$, $\bar{\gamma}$ and $\lambda$ all go to zero.  This may be shown via the definition of $\mathcal{A}_1$ and equations \ref{eqn:A1vlam} and \ref{eqn:gsimpvlam}. In this limit we can approximate $\mathcal{A}_1$ and $\bar{\gamma}$ by Taylor expanding equations \ref{eqn:A1vlam} and \ref{eqn:gsimpvlam}
\begin{eqnarray}
\mathcal{A}_1 &=& \frac{\lambda^2}{3} + O(\lambda^4) 
\label{eqn:Alamaprx},\\
\bar{\gamma}&=& \lambda^2 + O(\lambda^4) 
\label{eqn:gAapprox}.
\end{eqnarray}

\subsection{Axial pressure profile} \label{sec:compare}
We first compare the pressure profile along the axial direction with several other studies in the literature.  We note that in the limit of small permeability, $\lambda\rightarrow0$, and thus the pressure profile appears to be constant in $y$ as the leading order of $\cos(\lambda y)$ is 1.  This agrees with the results of\citet{Regirer60,Karode2001,Haldenwang2007,Tilton2012} to leading order.  Each of these works shows a similar profile in $x$ to the pressure profile we have presented, namely a linear combination of a  hyperbolic sine and cosine.  The scale of the arguments of these hyperbolic trig functions presented in \citet{Karode2001} and \citet{Tilton2012} is $\sqrt{3 \mathcal{A}_1}$ which we have shown is roughly $\lambda$ in the limit of small permeability (equation \ref{eqn:Alamaprx}).  The non-dimensional approximation for the pressure profile is in agreement in \citet{Karode2001} and \citet{Tilton2012} and is given as
\begin{align}
P(x,y) = \sqrt{3 \mathcal{A}_1} \sinh(\sqrt{3 \mathcal{A}_1} x) +  3 \mathcal{A}_1 \mathcal{B}_1 \cosh(\sqrt{3 \mathcal{A}_1} x),
\label{eqn:approxp}
\end{align}
where we will use a capital $P$ to denote the asymptotic approximation.
In \citet{Regirer60}, a solution for pipe rather than channel flow is performed, however we have confirmed that an identical result is found when the methodology is applied to channel flow.  To see this quickly we note that \citet{Regirer60} performs an asymptotic expansion about small Reynolds number in which the permeability is assumed to be small and we discuss this further below.  We also note that \citet{Haldenwang2007} develops the solution of \citet{Regirer60} before extending the work to non-zero transverse Reynolds number and therefore we have a similar result for all four works.

\begin{figure}
\includegraphics[height=6cm]{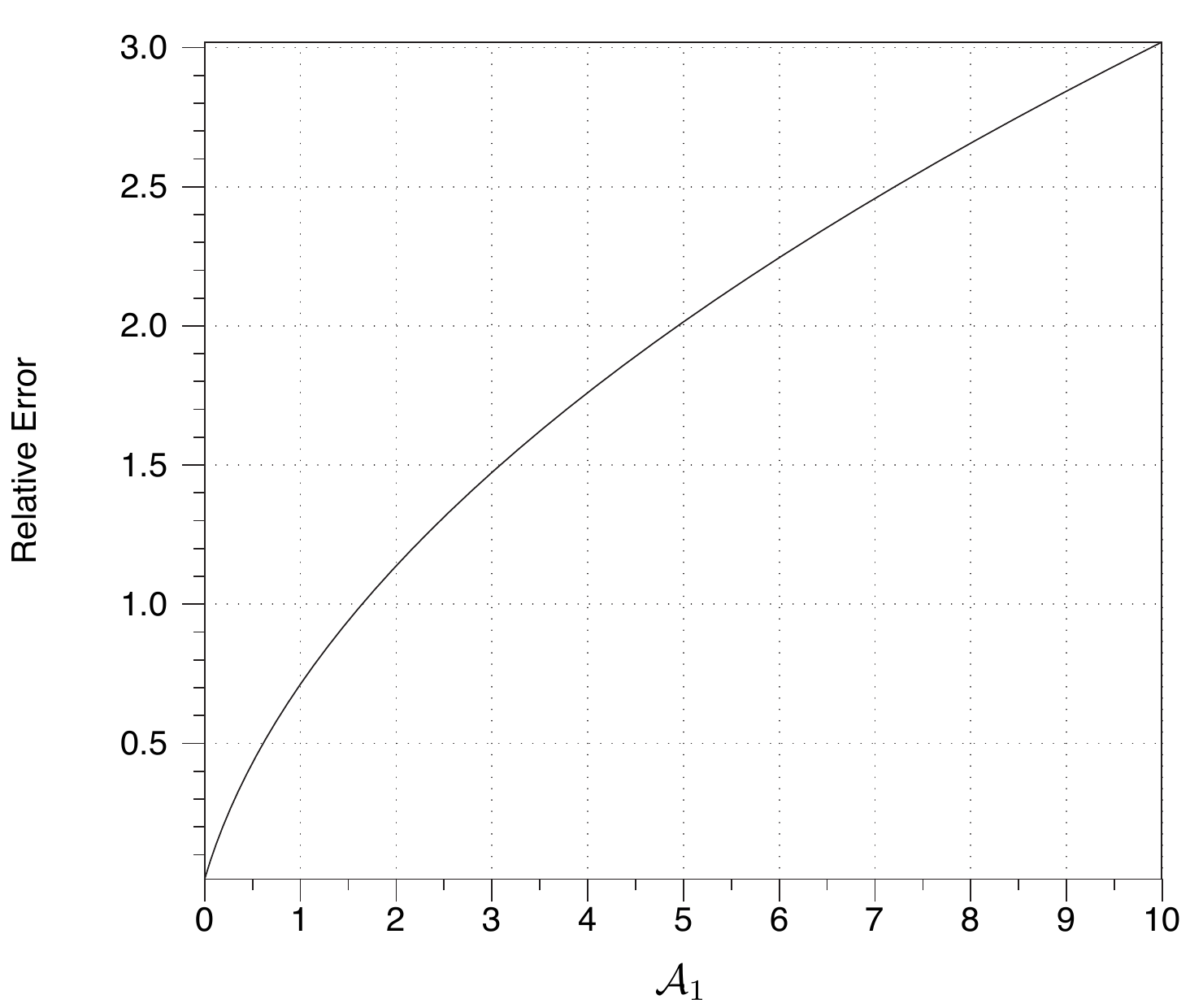}
\caption{We plot the relative error between the exponential scaling $\lambda$ and $\sqrt{3 \mathcal{A}_1}$.  The latter result is a prediction of the asymptotic theory and is seen to be the first order term in a Taylor expansion for $\lambda$ about $\lambda=0$.}
\label{fig:pprofileinc}
\end{figure}

Equation \ref{eqn:approxp} can be shown to approximate equation \ref{eqn:psol} in the limit of small permeability by applying \ref{eqn:Alamaprx} and Taylor expanding $\tan(\lambda)$ and $\cos(\lambda)$ about $\lambda=0$.  To assess the extent to which the prediction for axial pressure increase is approximated in the limit of small permeability (given that we will have exponential increase like $\exp(\lambda x)$ rather than $\exp(\sqrt{3 \mathcal{A}_1} x)$), we examine the difference between $\lambda$ and $\sqrt{3 \mathcal{A}_1}$.  We plot the relative error
\begin{align}
\frac{|\lambda-\sqrt{3 \mathcal{A}_1}|}{\lambda},
\end{align}
in figure \ref{fig:pprofileinc}.  We numerically determine the linear behavior of the relative error about $\mathcal{A}_1=0$ via a linear regression, and find that for small $\mathcal{A}_1$ the relative error is roughly $1.124 \mathcal{A}_1$.  The error in the asymptotic limit for large $x$ may then be approximated as the difference between exponentials and thus the relative error between the asymptotic approximation and the actual solution at zero Reynolds number is given as 
\begin{align}
\frac{|p(x,y)-P(x,y)|}{|p(x,y)|}\leq C_1 |1-\exp(-1.124 \mathcal{A}_1 x)|
\end{align}
for $x$ large and $\mathcal{A}_1$ small.
This estimate is novel.  

We note that there is a great deal of interest in the length of a pipe over which a solution may be valid.  Our result suggests that the existing asymptotic expansion  about small Reynolds number and our theory will only agree over axial domains that have length significantly less than $(-1.124 \mathcal{A}_1)^{-1}$ in non-dimensional units, however we note the caveat that the Reynolds number may not continue to be small within this regime.  The magnitude of the velocity in $v$ and $u$ along the axial direction is directly related to that of the pressure, and thus a similar statement may be made for the relative error in these profiles.  

In the present solution, the fields grow asymptotically with increasing $x$ like $C\exp(\lambda x)$ which will give a new estimate for the neighborhood of $x$ for which the velocity profile is below some threshold, thus ensuring convective terms may be ignored.  Note that $\lambda$ is a convex function of $\mathcal{A}_1$ and that $\lambda<\sqrt{3 \mathcal{A}_1}$ for all $\lambda$.  This means that the neighborhood of validity in $x$ for Stokes flow, where convective terms will not be important, is larger than the neighborhood predicted by the previous theory which predicts a faster asymptotic growth rate of the profiles in axial given by $C\exp(\sqrt{3 \mathcal{A}_1} x)$ .

\subsection{Transverse velocity profile}
We note that in the previous works, leading order behavior in the profiles for $u(0,x)$ and $v(0,x)$ are given by parabolic and polynomial expressions, respectively.
The solution for $u$ described in equation \ref{eqn:sol1i} may also be seen to be parabolic in $y$ to the leading order.  This may be achieved by Taylor expanding each term in the infinite sum about $\lambda=0$ and noting that to the leading order ($O(\lambda^0)$) the coefficients are a Fourier expansion of $(1-y^2)$ in the $y$ coordinate.  Similarly we may find that the leading order profile in $v$ depends on $y$ as $(y^3-3y)$ which may be found by noting that the two terms (the sum and $\sin(\lambda y)$) may be found in the leading order to be 
\begin{align}
v(x,y) \approx G(x) \lambda (y-y^3) + 2 G(x) \lambda y = \lambda (3 y - y^3),
\end{align}
where $G(x)$ is some function of $x$.
We further remark that this is precisely the leading order expression found in  \citet{Berman1953} and \citet{Tilton2012}.

Next we seek to better understand the asymptotic approximation breaks down in the transverse velocity profiles $u(\cdot,y)$ and $v(\cdot,y)$.  We note that both the solution presented in the present work and the solutions to the asymptotic approximations in the literature provide self similar profiles so that we may concern ourselves purely with the prediction of the shape in the profile.

\begin{figure}
\begin{center}
\includegraphics[height=5cm]{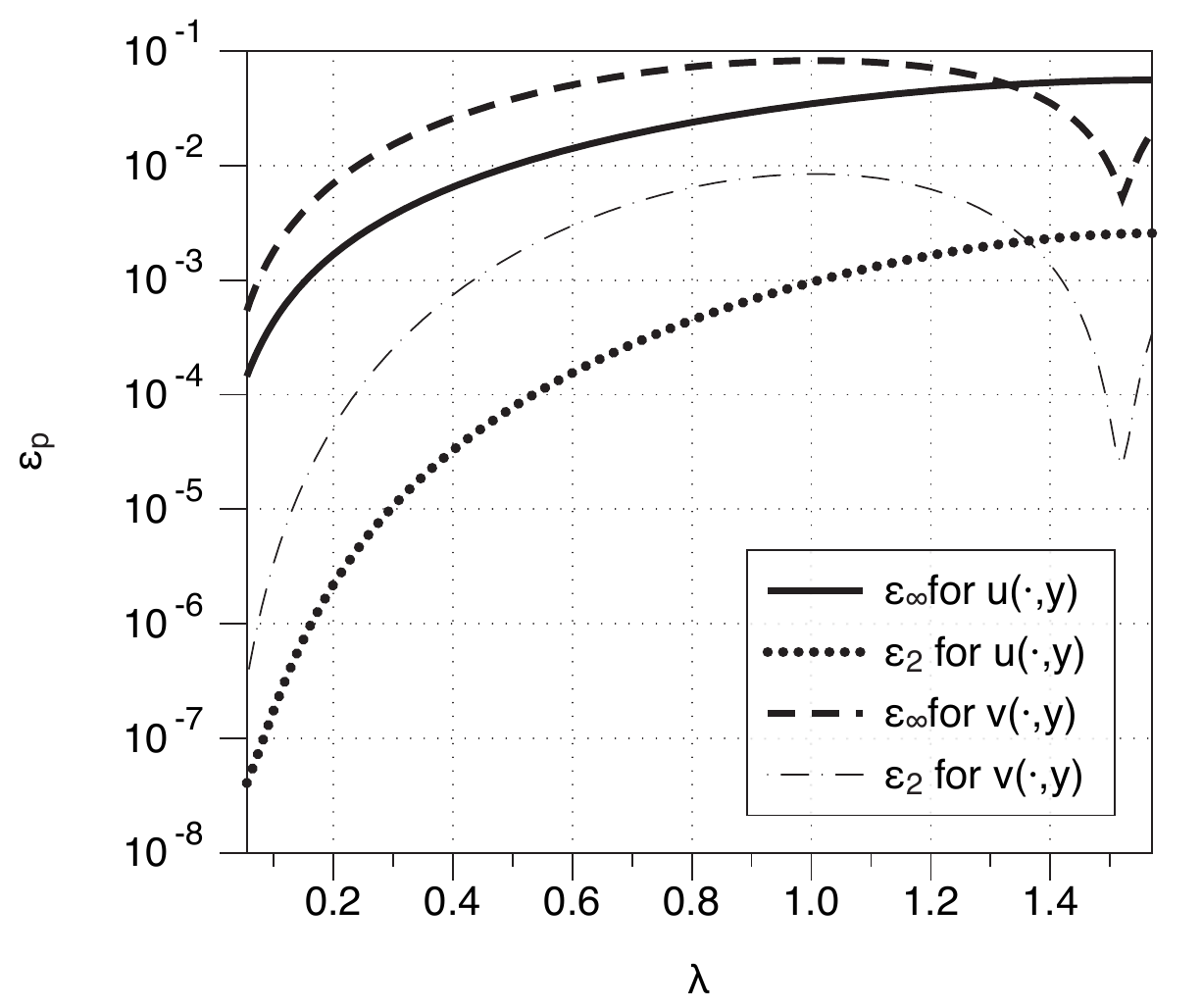}\\
\includegraphics[height=5cm]{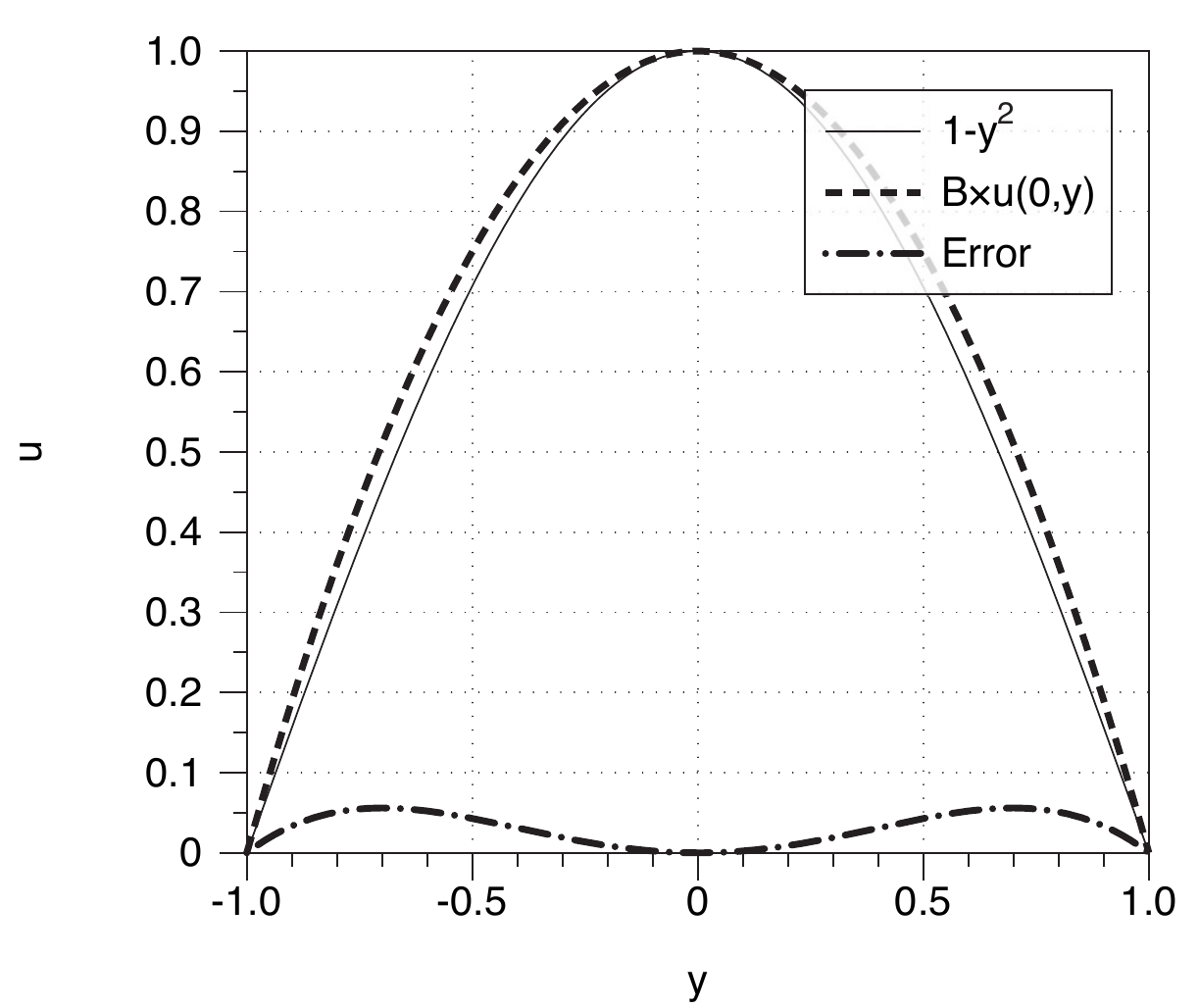}\\
\end{center}
\caption{The relative error defined by equation \ref{eqn:errsimp} is plotted as a function of $\lambda$ for the transverse profiles of $u$ and $v$ (top).  The flow profiles in $u$ are compared between parabolic flow and the present work for large permeability (bottom).  The solid dashed line corresponds to the high permeability profile where we have set $\lambda=0.9999\times\pi/2$, or, equivalently, $\mathcal{A}_1 = 2\times10^7$.  The error is plotted for comparison as the dot-dash line.}
\label{fig:relerrandvis}
\end{figure}

We begin with the profile in $u$ which is proportional to the profile
\begin{eqnarray}
u(\cdot,y) &\propto& \sum_{n=0}^\infty \frac{(-1)^{n+1}(1+2n)}{\left((1+2n)^2-\left(\frac{2 \lambda}{\pi}\right)^2\right)^2} \cos\left(\frac{\pi y}{2}+\pi n y\right).
\end{eqnarray}
We may then examine the relative error between our predicted profile and a parabolic profile global error in the following sense.  First, we set $B = 1/u(0,0)$ and scale $u$ by $B$ so that the profiles agree at $y=0$.  Next, we calculate the relative error between $(1-y^2)$ and $B\times u(0,y)$ via
\begin{eqnarray}
\epsilon_p(\lambda) \equiv \frac{||(1-y^2)-B\times u(0,y)||_p}{||B\times u(0,y)||_p},
\label{eqn:errsimp}
\end{eqnarray}
where $||\cdot||_p$ is the $L^p([-1,1])$ norm; we consider the error for $p\in\{2,\infty\}$ (see figure \ref{fig:relerrandvis}).  In the $L^\infty$ and $L^2$ norms we find that the relative error increases to over 5\% and 0.3\% in the limit of large permeability, respectively (see figure \ref{fig:relerrandvis}).  We also note that our error bounds could be even tighter if we redefine the relative error to be 
\begin{equation}
\epsilon_p^l(\lambda) \equiv \inf_{B} \frac{||(1-y^2)-B\times u(0,y)||_p}{||B\times u(0,y)||_p}.
\end{equation}
The result is that we have shown how the existing theory for small permeability breaks down once permeability can no longer considered small.  Although the error is small, flow profile is not parabolic for non-zero permeability.

Next we wish to examine the profile in $v$
by analyzing the self similar profile $v(\cdot,y)$.  In order to show the difference between the asymptotic approximation and our solution, we again consider the relative error between the asymptotic profile and our profile, rescaling so that the velocity at the channel wall, described by the Darcy condition, is equivalent in each case; this is to say we compare $(3 y-y^3)/2$ with $v(\cdot,y)/v(\cdot,1)$.  We then analyze the relative error and plot the results in figure \ref{fig:relerrandvis}.  We find that the error achieves a maximum not in the limit of large permeability but at an intermediate permeability corresponding to $\lambda = 1.0105$ or $\mathcal{A}_1=0.98$.  The maximum relative error is 8\% and 0.8\% respectively in the infinity and two norms which is larger than the error in the profile for $u$.

\subsection{Axial flow exhaustion and crossflow reversal}
\begin{figure}
\begin{center}
\includegraphics[height=5cm]{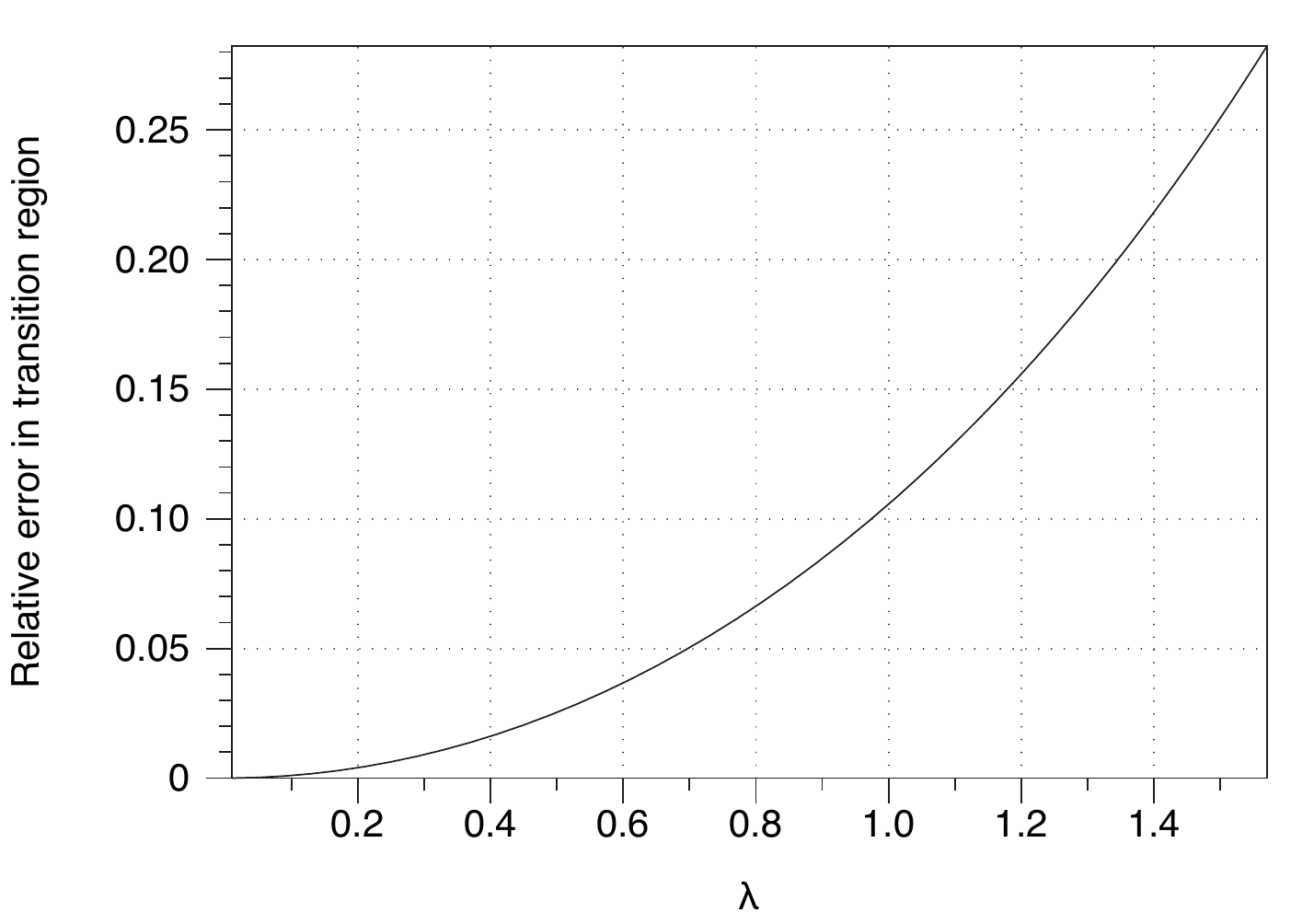}\\
\includegraphics[height=5cm]{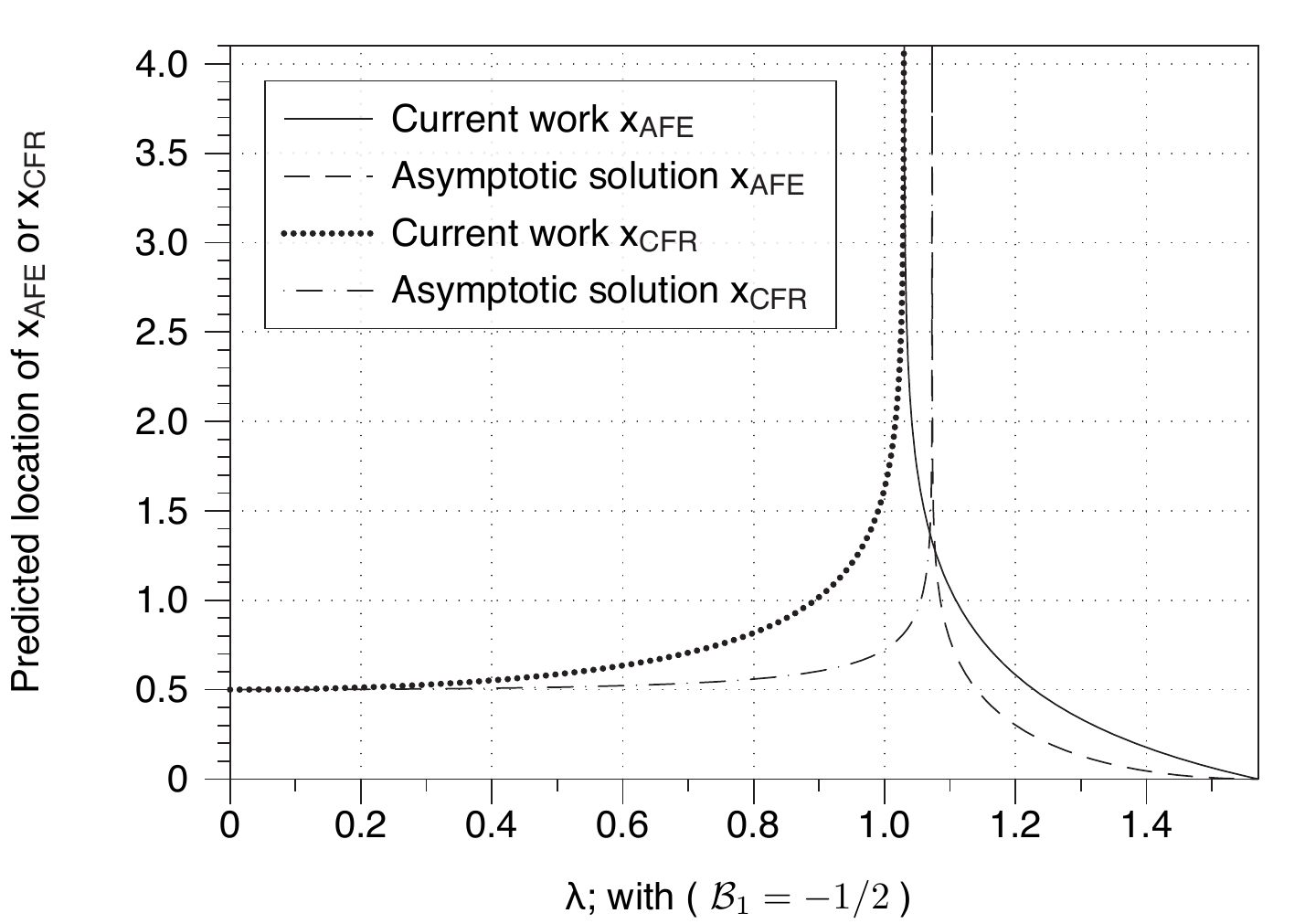}\\
\end{center}
\caption{The relative error in the predicted transition value for $\mathcal{B}_1$ between axial flow exhaustion and no axial flow exhaustion is plotted as a function of $\lambda$ (top).  Next, we fix $\mathcal{B}_1=-1/2$ and show the predicted location of $x_{AFE}$ as a function of $\lambda$ (bottom).}
\label{fig:cf}
\end{figure}

Two important flow behaviors that occur in channels and pipes with permeable walls are axial flow exhaustion and crossflow reversal.  Axial flow exhaustion occurs when the axial velocity profile ($u$) changes sign in the axial direction (i.e. at some position $x$).  This occurs because flow is either being injected or suctioned out of the channel with high enough pressure difference (i.e. large $\mathcal{B}_1$) to break down axial flow through the channel at a point.  This corresponds to a position in $x$, denoted $x_{AFE}$, such that $u(x_{AFE},y) = 0$.  Crossflow reversal is described by a change of sign in the normal velocity along the channel wall. Physically this is described by the channel walls transitioning from fluid suction to fluid injection or vice-versa, and occurs when $\mathcal{B}_1$ is small.  The axial position where cross flow reversal occurs will be denoted $x_{CFR}$ and at this position $v(x_{CFR},y)=0$.  In this section we compare the predictions of where, and whether axial flow exhaustion or crossflow reversal occurs by comparing the leading order asymptotic analysis of \citet{Tilton2012} with our solution.  By setting $u(x_{AFE},y)=0$ we may solve to find the following two predictions for the asymptotic theory and the present work respectively:
\begin{eqnarray}
&x_{AFE} &= \frac{1}{\sqrt{3\mathcal{A}_1}}\arctanh\left(-\frac{1}{\mathcal{B}_1 \sqrt{3\mathcal{A}_1}}\right),\\
&x_{AFE} &= \frac{1}{\lambda}\arctanh\left(-\frac{\cos(\lambda)}{\mathcal{B}_1 \lambda}\right).
\end{eqnarray}
%
We can also determine the location of cross flow reversal by setting $v(x_{CFR},y)=0$ and find the following two predictions for the asymptotic theory and the present work respectively:
\begin{eqnarray}
x_{CFR} &= \frac{1}{\sqrt{3\mathcal{A}_1}}\arctanh\left(-\mathcal{B}_1 \sqrt{3\mathcal{A}_1}\right),\\
x_{CFR} &= \frac{1}{\lambda}\arctanh\left(-\frac{\mathcal{B}_1 \lambda}{\cos(\lambda)}\right).
\end{eqnarray}
As is noted in \citet{Haldenwang2007}, the two regimes are mutually exclusive.  This can be seen by noting that the arguments of the $\arctanh$ function for $x_{AFE}$ and $x_{CFR}$ are reciprocals in both theories; that is when $x_{AFE}$ is real, $x_{CFR}$ is imaginary and vice-versa.  We may then ask where in the parameter space does the solution transition from exhibiting cross flow reversal to axial flow exhaustion.  This will occur when the argument of the $\arctanh$ function is $\pm 1$, or rather when
\begin{eqnarray}
\left|\frac{1}{\mathcal{B}_1 \sqrt{3\mathcal{A}_1}}\right|=1, \quad \left|\frac{\cos(\lambda)}{\mathcal{B}_1 \lambda}\right|=1
\label{eqn:bfcond}
\end{eqnarray}
in the respective theories.  We can then examine the relative error in the predicted transition between axial flow exhaustion and crossflow reversal within the parameter space.  To do this we analyze the predicted transition value for $\mathcal{B}_1$ as a function of $\lambda$ which we can define in the respective theories as
\begin{eqnarray}
\mathcal{B}_1^T = \pm\left|\frac{1}{ \sqrt{3\mathcal{A}_1}}\right|, \quad \mathcal{B}_1^T=\pm\left|\frac{\cos(\lambda)}{ \lambda}\right|=1.
\label{eqn:bfcond}
\end{eqnarray}
We define the relative error in the two predicted transition values for $\mathcal{B}_1$ as
\begin{equation}
\frac{|\cos(\lambda)/\lambda-\sqrt{3 \mathcal{A}_1}|}{|\cos(\lambda)/\lambda|},
\end{equation}
which we display in figure \ref{fig:cf}.
We find that the relative error for the cut off condition on $\mathcal{B}_1$ grows with $\lambda$ and that for permeability the error grows to be over 25\%.  
Finally we examine the error in the prediction for the location of $x_{AFE}$ and $x_{CFR}$ at a fixed value of $\mathcal{B}_1$.  We note that as $\mathcal{B}_1$ approaches the transition value, that the location of $x_{AFE}$ and $x_{CFR}$ goes to infinity in both the asymptotic and our prediction. 
To make a demonstrative comparison we set $\mathcal{B}_1=-1/2$ and compare the predictions of $x_{AFE}$ and $x_{CFR}$ in figure \ref{fig:cf}.  We find that for large values of $\lambda$ the error is of order 1 near the transition regions which further demonstrates the break down of the asymptotic prediction.  We remark that when $\lambda$ is small we expect that $\mathcal{B}_1$ must be large in order for axial flow exhaustion to occur which corresponds either to large values of transmembrane pressure $P_{tm}$ or small values of the pressure gradient $\beta$ or, equivalently, average axial velocity profile $\bar{U}$.

\section{Numerical experiment comparison for short channels with convective effects}
\label{sec:numerical}
It has been shown that as permeability increases, the convective terms of the Navier-Stokes equations become more important at shorter axial length scales (see refs. \onlinecite{Haldenwang2007,Tilton2012,Bernales2014}). We will assess the extent to which the convection terms affect the accuracy of our solution and determine parameter regimes for which our solution is more accurate than the existing asymptotic theory of \citet{Tilton2012}.

The Navier-Stokes equations, under the assumption of laminar flow, are nondimensionalized as above with $\alpha=1$ and the spatial variables are rescaled so that the half width of the channel is one.   The resulting system of equations is
\begin{eqnarray}
&&p(x_0,\pm 1) = \mathcal{B}_1, \quad  p_x(x_0,0)=\gamma,\\
&& Re_c(uu_x+vu_y)= -p_x+\mathcal{A}_1 \Delta u,\label{eqn:NSu}\\
&& Re_c(uv_x+vv_y)= -p_y+\mathcal{A}_1 \Delta v,\label{eqn:NSv}\\
&&0=u_x+v_y,\\
&&u(x,\pm 1) = 0,\quad v(x,\pm 1) = \pm p,
\end{eqnarray}
where $Re_c=\rho \kappa^2 \beta r/\gamma$, and $\rho$ is the density of the fluid.  The non-dimensional constant $Re_c$ is a rescaled value of the Reynolds number, where the Reynolds number is defined as $Re = \rho \bar{U} 2 r/\mu$. There is a one-to-one correspondence between $Re$ and $Re_c$.  Utilizing equation \ref{eqn:onetoonebetaubar}, we obtain the linear scaling
\begin{align}
Re = \frac{Re_c}{\mathcal{A}_1} \frac{\int_{-r}^r f(y;\lambda)}{r} dy.
\end{align}
For the remainder of this section we will present our results in terms of $Re$, rather than $Re_c$ to remain consistent with previous work (e.g. ref. \onlinecite{Tilton2012}).
We also note that we may use the Stokes flow solution of the current work to be the zeroth order term of an asymptotic expansion about low Reynolds number as is done by \citet{Regirer60}, and \citet{Bernales2014}.  Our solution is different than these pervious works in that we do not assume that the permeability is negligible in the zeroth order term.

\begin{figure}
\begin{center}
\includegraphics[height=7.5cm]{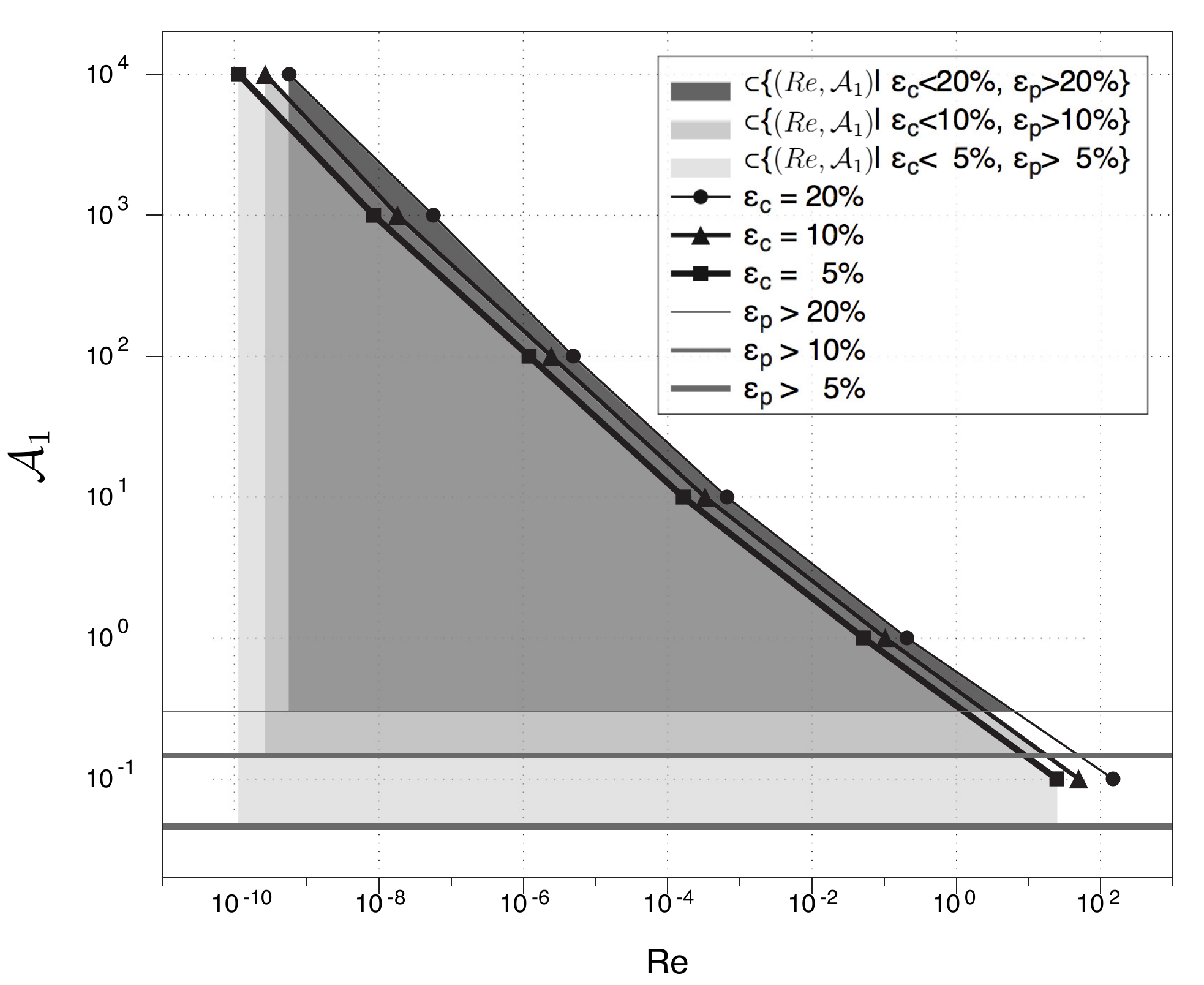}\\
\end{center}
\caption{Level set curves of $\epsilon_c$ equal to 5\%, 10\% and 20\% relative errors are presented for the present work when compared to numerical solutions of the Navier-Stokes equations (black lines with circles, diamonds, and squares, respectively).   We display grey lines, above which $\epsilon_p$ is greater than 5\%, 10\% and 20\% (grey lines in decreasing thickness). We display regions in which the current work outperforms the pervious work in the parameter space with light grey (5\%), grey (10\%), and dark grey (20\%) colorations.}
\label{fig:numericalresults}
\end{figure}

It is reasonable to conjecture that our solution will be more accurate than the existing theory at high permeability and bounded channel length.  The idea is that pervious theory will break down as permeability increases, whereas the present theory will not provided that the left hand sides of equations \ref{eqn:NSu} and \ref{eqn:NSv} remain relatively small.  Due to the exponential increase in the velocity profiles, the left hand sides of equations \ref{eqn:NSu} and \ref{eqn:NSv} will only remain small so long as either the channel length is small or $Re_c$ is small.   Increasing the length of the channel will require smaller values for $Re_c$ for the present work to accurately approximate solutions to the Navier-Stokes equations.  Despite the requirement that  $Re_c$ is small, we still expect to find regions in the parameter space in which the current solution outperforms the existing theory at high permeability, independent of the channel size.  Because we are interested solely in numerically validating the above ideas, we examine a channel with small length, setting the length to be five times the half width.
We then analyze the parameter space by examining the parameter values $\log_{10}(\mathcal{A}_1) = \{i\}_{i=-1}^4$ and $\mathcal{B}_1=1$.  The code is run on a MAC grid and we use the Newton-Krylov method found in the ``optimize" package of scipy.  Inlet and outlet conditions are assumed to agree with the present theory (see equations \ref{eqn:psol}-\ref{eqn:sol1f}). Spatial resolution is taken to be $2^6$ divisions per unit of space on a MAC grid.  This resolution was chosen because the relative error between the analytic solution found in the present work and the numerical result at Stokes flow was bounded by a relative error of less than $5\times10^{-4}$ over all permeabilities.  At each non-dimensional permeability we increase the Reynolds number to determine the relative error as a function of $Re$ and $\mathcal{A}_1$ which is defined as
\begin{align}
\epsilon_{j}(Re,\mathcal{A}_1)\equiv \max\left(err_2, err_\infty \right),
\end{align}
where $j\in\{c,p\}$ for maximum relatives errors between numerical simulations found between the ``current" and ``previous" work respectively,
\begin{align}
err_i = \max\left(\frac{||u_T-u_N||_i}{||u_T||_i},\frac{||v_T-v_N||_i}{||v_T||_i},\frac{||p_T-p_N||_i}{||p_T||_i}\right),
\end{align}
$u_T,v_T,p_T$ represent field variables determined from theory and $u_N,v_N,p_N$ represent field variables determined from numerical experiment.  In figure \ref{fig:numericalresults}, we plot the  5\%, 10\% and 20\%  level set curves of $\epsilon_{c}(Re,\mathcal{A}_1)$.  Below and to the left of each threshold line, the relative error between the Navier-Stokes numerical solution and the predictions from the current work are below the relative error threshold.

We would like next to determine subregions where our work outperforms the existing theory of \citet{Tilton2012}.  A natural way to do this is to repeat the above numerical experiment using inlet/outlet conditions that are consistent with the solution of ref. \onlinecite{Tilton2012} so that we do not bias our result, and then determine regions in the parameter space for which the relative error is \emph{below} $x$ percent for the current work and \emph{above} $x$ precent for the previous work.   However such an experiment would double the computational cost of the numerical scheme listed above, and this scheme is already expensive since we are sweeping over many values in the $(Re,\mathcal{A}_1)$ parameter space.  

To avoid this large expense, we note instead that the work of \citet{Tilton2012} presents an asymptotic expansion about low Reynolds number.  This means that the maximum relative error at zero Reynolds number is less than the maximum relative error at non-zero Reynolds number for fixed $\mathcal{A}_1$, which is to say
\begin{align}
\epsilon_{p}(0,\mathcal{A}_1) \leq \epsilon_{p}(Re,\mathcal{A}_1), \quad \forall \mathcal{A}_1, Re.
\end{align}
With this observation, and noting again that we are only interested in demonstrating the existence of regions with enhanced performance in the present theory, we determine $\epsilon_{p}(0,\mathcal{A}_1)$ for the values of $\mathcal{A}_1$ listed above ($\log_{10}(\mathcal{A}_1) = \{i\}_{i=-1}^4$).  We then determine the values of $\mathcal{A}_1$ for which $\epsilon_{p}(0,\mathcal{A}_1)$ is equal $5\%, 10\%$ and $20\%$.  The intersection of the regions where $\epsilon_{p}(0,\mathcal{A}_1)>x$ and where $\epsilon_c<x$ is a region in which the present work outperforms the previous work.  We plot the regions in figure \ref{fig:numericalresults} for $x\in\{5\%,10\%,20\%\}$ and in all cases find non-empty regions where the current work outperforms the previous theory.  We have examined the numerical results for other values of $\mathcal{B}_1$ and find qualitatively similar results. 


\section{Discussion}
\label{sec:disc}

We have considered Stokes flow through an infinite channel with permeable walls, such that fluid may be driven by pressure differences across the channel wall to enter or exit the channel.  The normal flux component is given by Darcy's law, whereas the tangential component is assumed to be zero (i.e., no-slip boundary condition).  The novel element to our work is that the permeability may be of arbitrary magnitude.  Such a solution allows us to directly and analytically test the break down of existing asymptotic theories at small Reynolds number and we have demonstrated how the theory breaks down in terms of the axial magnitude of the pressure, the transverse velocity profiles, and as a predictive tool for axial flow exhaustion and crossflow reversal.  We note that our solution is equivalent to examining the case in which transverse velocity is no longer negligible when compared to the axial velocity profile.  We have also contextualized our work in terms of an asymptotic expansion about small Reynolds number.  Although we have not attempted to determine the higher order corrections, we note that if such corrections could be made, the theory of \citet{Regirer60} would be extended to a far broader region of the non-dimensionalized parameter space.

In addition, our solution also extends the known analytic results for a wider class of parameters than has been originally explored.  The utility of such an exploration is, as of yet, to be seen, as most values for $\mathcal{A}_1$ found in nature may be considered to be significantly less than one.  We do however note that our result may be useful in setting inlet and outlet boundary conditions for numerical studies.  In the numerical study by \citet{Pozrikidis2010}, the author studies fluid loss in capillaries.  Boundary conditions are set by proposing a parabolic profile and then allowing a region of pipe with zero permeability to transition into a region with non-zero permeability.  It was shown that the pressure profile spikes upon this transition and thus in a study involving the concentration of a secondary passive scalar we may see unphysical loss at these regions.  Furthermore, there is added computational expense involved in setting these boundary conditions, as there is a required transition region from permeable to impermeable wall.  As a potential remedy, our flow profiles may act as inlet/outlet profiles that do not cause unrealistic spikes in the pressure profiles and may also reduce the number of grid points needed at a boundary.

\begin{figure}
\includegraphics[width=6cm, clip=true, trim=11cm 0cm 11cm 0cm]{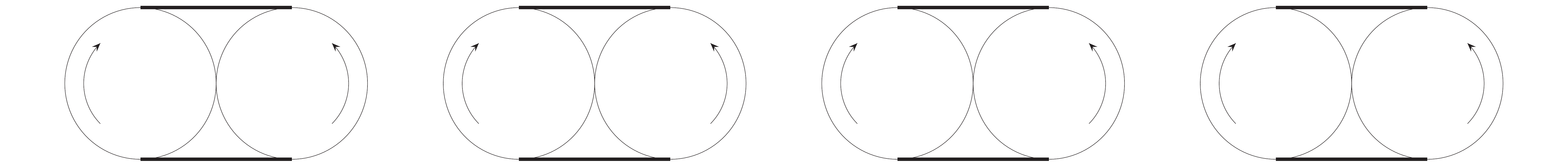}
\caption{Theoretical microscopic detail of a membrane with potentially large permeability. Cylinders rotate with passing fluid and act as lubricating elements to avoid the issue of no slip condition though the microscopic permeable pathways linking the exterior and interior of the channel.}
\label{fig:theofyflow}
\end{figure}

In practice, the non-dimensional permeability is typically small, however we propose a theoretical framework in which it may be considered to be arbitrarily large.  The idea is to imagine the channel wall to be a series of coupled cylinders with small spacing as pictured in figure \ref{fig:theofyflow}.  A pressure difference across the channel would then cause flow across the membrane which would rotate the cylinders; a frictional coefficient on the cylinders would determine the relationship between the pressure and the outward velocity that would be significantly larger than if the cylinders were fixed due to the fact that we circumvent the frictional restrictions of the zero slip condition.  Such a physical set up may allow for novel filtration applications in engineering where one desires filtration within a small channel.

\section*{Acknowledgements}

This research was supported by the National Institutes of Health, National Institute of Diabetes and Digestive and Kidney Diseases, grant DK089066, to A. Layton; by the National Science Foundation, grant DMS1263995, to A. Layton, Research Training Groups grant DMS0943760 to the Mathematics Department at Duke University, Research Network in the Mathematical Sciences grant DMS1107444 to KI-Net, and NSF grant DMS1514826 to Jian-Guo Liu.  We would like to acknowledge Thomas P. Witelski for fruitful discussions and generously editing and critiquing our manuscript.

\appendix
\section{}\label{appA}
\begin{proposition}\label{prop:yind}
The quantity 
\begin{equation}
\sum_{n=0}^\infty\left(\frac{d_n}{\omega_n^2-\lambda_0^2} \frac{\cos(\omega_n y)}{\cos(\lambda_0 y)}-\frac{\bar{d}_n \bar{\omega}_n}{\lambda_0 (\bar{\omega}_n^2-\lambda_0^2)} \frac{\cos(\bar{\omega}_n y)}{\cos(\lambda_0 y)}\right),
\label{eqn:yindhuh}
\end{equation}
from equation \ref{eqn:alphasimp}, is independent of $y$ for all values of $\lambda_0\in(0,\pi/2)$, given by equation \ref{eqn:eigbound}.  
\end{proposition}

To prove proposition \ref{prop:yind}, we first let
\begin{equation}
F(\lambda_0,y) =\sum_{n=0}^\infty\left(\frac{d_n}{\omega_n^2-\lambda_0^2} \cos(\omega_n y)-\frac{\bar{d}_n \bar{\omega}_n}{\lambda_0 (\bar{\omega}_n^2-\lambda_0^2)} \cos(\bar{\omega}_n y)\right),
\end{equation}
We then claim that $F(\lambda_0,y)=C(\lambda_0)\cos(\lambda_0 y)$ which will demonstrate the loss of $y$ dependence.  To show this we demonstrate the equivalence of the Fourier modes by finding the correct scaling $C(\lambda_0)$.  This is equivalent to showing
\begin{equation}
\left\langle\cos(k\pi y),F(\lambda_0,y)\right\rangle = C(\lambda_0) \left\langle\cos(k\pi y),\cos(\lambda_0 y)\right\rangle,
\label{eqn:neccond}
\end{equation}
for $k\in\mathbb{N}$, with the inner product defined as usual to be
\begin{equation}
\left\langle f(y), g(y) \right\rangle = \int_{-\gamma}^\gamma  f(y) g(y) dy.
\end{equation}

For equation \ref{eqn:neccond} to be true, the $k=0$ case implies that we must have
\begin{equation}
\sum_{n=0}^\infty \frac{32 \gamma^2 \cos(\lambda_0\gamma)}{\pi^4\left((1+2k)^2-\left(\frac{2\lambda_0\gamma}{\pi}\right)^2\right)^2} = C(\lambda_0) \frac{\sin(\lambda_0\gamma)}{\lambda_0\gamma},
\end{equation}
which we have derived by integrating equation \ref{eqn:neccond}.  We simplify the left hand side via the identity
\begin{equation}
\sum_{n=0}^\infty \frac{1}{\left((1+2k)^2-x^2\right)^2} = \frac{\pi^2\sec^2\left(\frac{\pi x}{2}\right)}{16 x^2}- \frac{\pi\tan\left(\frac{\pi x}{2}\right)}{8 x^3},
\end{equation}
and then solve for $C(\lambda_0)$, which leads to the condition
\begin{equation}
C(\lambda_0) = \frac{1}{2}\left(\frac{\gamma}{\sin(\lambda_0\gamma)\cos(\lambda_0\gamma)\lambda_0}-\frac{1}{\lambda_0^2}\right).
\label{eqn:clamb}
\end{equation}
For $k>0$ we are left to verify that
\begin{widetext}
\begin{eqnarray}
\frac{ 2(-1)^{k} k^2 \pi^2 \gamma \sin(\lambda_0\gamma) }{\lambda_0(k^2 \pi^2-\gamma^2\lambda_0^2)^2} 
&+&\sum_{n=0}^\infty\frac{64 (-1)^k (1+2 n)^2 \gamma^2 \cos(\lambda_0\gamma)}{\pi^4\left((1+2 n)^2-4 k^2\right) \left((1+2 n )^2-\left(\frac{2 \lambda_0\gamma}{\pi}\right)^2\right)^2}
=  C(\lambda_0)\frac{2 (-1)^k \lambda_0\gamma \sin(\lambda_0\gamma )}{-k^2 \pi ^2+\lambda_0^2\gamma ^2}.
\end{eqnarray}
\end{widetext}
where we have again used \ref{eqn:neccond} to derive this formula.  The sum on the left hand side may be reduced via the identity
\begin{widetext}
\begin{equation}
\frac{\frac{\pi ^2 \lambda}{2} \left(4 k^2- \lambda^2\right) \sec\left(\frac{\pi \lambda}{2}\right)^2+\pi \left(4 k^2 +\lambda^2\right) \tan\left(\frac{\pi \lambda}{2}\right)}{8 \lambda\left(\lambda^2-4 k^2 \right)^2} = \sum_{n=0}^\infty\frac{(1+2 n)^2}{\left(4 k^2-(1+2 n)^2\right)\left((1+2 n)^2-\lambda^2\right)^2}.
\end{equation}
\end{widetext}
by substituting $\lambda=\frac{2\lambda_0\gamma}{\pi}$.  This leads to an algebraic expression that we have verified to be valid, however we have omitted the details as it leads to a lengthy reduction.  This completes the proof of proposition \ref{prop:yind}.

\begin{proposition}\label{prop:posmono} 
The relationship between $\mathcal{A}$ and $\lambda_0$, given in equation \ref{eqn:Avslam},
\begin{equation}
f(\lambda_0) = \frac{1}{2}\left(\frac{\gamma}{\cos^2(\lambda_0\gamma)}-\frac{\sin(\lambda_0\gamma)}{\lambda_0\cos(\lambda_0\gamma)}\right),
\end{equation}
with $f:(0,\pi(2\gamma)^{-1})\rightarrow(0,\infty)$ is bijective.
\end{proposition}

To prove this we first note that the function $f$ is continuous in $\lambda_0$.  We then need to show that
\begin{eqnarray}
\lim_{\lambda_0\rightarrow0} f(\lambda_0)&=&0,\\
\lim_{\lambda_0\rightarrow\pi(2\gamma)^{-1}} f(\lambda_0)&=&\infty,
\end{eqnarray}
and that $f$ is monotonically increasing.  The first limit can be seen by noting that $\lim_{\lambda_0\rightarrow0} \sin(\lambda_0 \gamma) = \lambda_0 \gamma$.  In the second limit, the first term of $f$ dominates and is unbounded from above.  To show that the function is monotonically increasing we first let $\lambda_0 \gamma=\lambda$ and then show that 
\begin{eqnarray}
\frac{df(\lambda)}{d\lambda} &> 0,
\end{eqnarray}
where
\begin{eqnarray}
\frac{2}{\gamma}\frac{df(\lambda)}{d\lambda} &=& 2\sec^2(\lambda)\tan(\lambda)\nonumber\\
&&¨+\tan(\lambda)\lambda^{-2}-\sec^2(\lambda)\lambda^{-1}
\end{eqnarray}
which will be true so long as
\begin{eqnarray}
2\sec^2(\lambda)\tan(\lambda)\lambda^2+\tan(\lambda)-\sec^2(\lambda)\lambda&>&0,\nonumber\\
2\tan(\lambda)\lambda^2+\sin(\lambda)\cos(\lambda)&>&\lambda.\label{eqn:hope1}
\end{eqnarray}
In the limit as $\lambda\rightarrow0$ is zero for both sides of equation \ref{eqn:hope1}.  Therefore it suffices to show that 
\begin{eqnarray}
\frac{d(2\tan(\lambda)\lambda^2+\sin(\lambda)\cos(\lambda))}{d\lambda}>\frac{d\lambda}{d\lambda},\nonumber\\
1-2\sin^2(\lambda)+4\lambda\tan(\lambda)+2\lambda^2>1,\nonumber\\
2\lambda\sin(\lambda)\cos(\lambda)+\lambda^2>\sin^2(\lambda)\cos^2(\lambda),
\end{eqnarray}
which is true since
\begin{eqnarray}
2\lambda\sin(\lambda)\cos(\lambda)+\lambda^2>\lambda^2, \\
 \lambda^2>\sin^2(\lambda)\cos^2(\lambda),
\end{eqnarray}
for $\lambda\in(0,\pi/2)$.  This completes the proof of proposition \ref{prop:posmono}.

%
%

\end{document}